\input harvmac

%%%%%%%%%%%%%%%%%%%%%%%%%%%%%%%%%%%%%%%%%%%%%%%%%%%%%%%%%%%%%%%%%%%
%%%  modify title page
%%%%%%%%%%%%%%%%%%%%%%%%%%%%%%%%%%%%%%%%%%%%%%%%%%%%%%%%%%%%%%%%%%%
\def\Title#1#2{\rightline{#1}\ifx\answ\bigans\nopagenumbers\pageno0
%   \vskip0.5in
\else\pageno1\vskip.5in\fi \centerline{\titlefont #2}\vskip .3in}

\font\caps=cmcsc10
%\def\listrefs{\footatend\bigskip\bigskip\immediate\closeout\rfile
%\writestoppt \baselineskip =13pt\centerline{{\secfont References}}
%\bigskip{\frenchspacing\parindent =20pt \escapechar +'
%\input\jobname.refs \vfill\eject}\nonfrenchspacing}
%%%%%%%%%%%%%%%%%%%%%%%%%%%%%%%%%%%%%%%%%%%%%%%%%%%%%%%%%%%%%%%%%%%%%%%%%%%%

\noblackbox
\parskip=1.5mm
%\def\semi{;~}

%%%%%%%%%%%%%%%%%%%%%%%%%%%%%%%%%%%%%%%%%%%%%%%%%%%%%%%%%%%%%%%%%%%%%

\def\npb#1#2#3{{\it Nucl. Phys.} {\bf B#1} (#2) #3 }
\def\plb#1#2#3{{\it Phys. Lett.} {\bf B#1} (#2) #3 }
\def\prd#1#2#3{{\it Phys. Rev. } {\bf D#1} (#2) #3 }

\def\mpla#1#2#3{{\it Mod. Phys. Lett.} {\bf A#1} (#2) #3 }

\def\bb#1{{\tt hep-th/#1}}

%%%%%%%%%%%%%%%%%%%%%%%%%%%%%%%%%%%%%%%%%%%%%%%%%%%%%%%%%%%%%%%%%%%%%
%%%%%%%%%%%%%%%%%%%%    some definitions    %%%%%%%%%%%%%%%%%%%%%%%%%
%%%%%%%%%%%%%%%%%%%%%%%%%%%%%%%%%%%%%%%%%%%%%%%%%%%%%%%%%%%%%%%%%%%%%

%%%%%%%%%%%%%%%%%%%%%%%%%%%%%%%%%%%%%%%%%%%%%%%%%%%%%%%%%%%%%%%%%%%%%

\def\dj{\hbox{d\kern-0.347em \vrule width 0.3em height 1.252ex depth
-1.21ex \kern 0.051em}}

%%%%%%%%%%%%%%%%%%%%%%%%%%%%%%%%%%%%%%%%%%%%%%%%%%%%%%%%%%%%%%%%%%%%%%
%%%%%%%%%%%%%%%%%%%%%%%        references         %%%%%%%%%%%%%%%%%%%%%%
%%%%%%%%%%%%%%%%%%%%%%%%%%%%%%%%%%%%%%%%%%%%%%%%%%%%%%%%%%%%%%%%%%%%%%%dsky

\lref\maldacena{J. Maldacena, {\it ``The Large $N$ Limit
 of Superconformal Field Theories and Supergravity'',}
 \bb{9711200.}}
\lref\wittenads{E. Witten, {\it
``Anti-de Sitter Space, Thermal Phase Transition, And
Confinement in Gauge Theories'',} \bb{9803131.}}
\lref\rwittentheta{E. Witten, {\it ``Theta dependence in the large $N$
limit of four-dimensional gauge theories'',} \bb{9807109.}}
\lref\givkut{A. Giveon and D. Kutasov, {\it ``Brane Dynamics and 
Gauge Theory'',} \bb{9802067.}}
\lref\rmberk{K. Hori, H. Ooguri and Y. Oz, 
{\it ``Strong Coupling Dynamics of Four-Dimensional $N=1$ 
Gauge Theories from M Theory Fivebrane",} 
{\it Adv. Theor. Math. Phys.} {\bf 1} (1998) 1,
\bb{9706082.}}
\lref\rwitqcd{E. Witten, 
{\it ``Branes And The Dynamics Of QCD",}
\npb{507}{1997}{658,}
\bb{9706109.}}
\lref\rthetus{J.L.F. Barb\'on and A. Pasquinucci, 
{\it ``Softly Broken MQCD and the Theta Angle'',}
\plb{421}{1998}{131,} 
\bb{9711030.}}
\lref\rbrokus{J.L.F. Barb\'on and A. Pasquinucci,
{\it ``A Note on Softly Broken MQCD'',}
\mpla{13}{1998}{1453,}
\bb{9804029.}}
\lref\rmasiero{A. Masiero and G. Veneziano, 
{\it `` Split Light Composite Supermultiplets'',}
\npb{249}{1985}{593.}}
\lref\VY{G. Veneziano and S. Yankielowicz,
{\it ``An Effective Lagrangian for the pure N=1 Supersymmetric Yang-Mills
theory'',}
\plb{113}{1982}{231.}} 
\lref\TVY{T. Taylor, G. Veneziano and S. Yankielowicz,
 {\it ``Supersymmetric QCD and its Massless Limit: An Effective Lagrangian
Analysis'',}
\npb{218}{1983}{493.}}
\lref\dhoo{J. de Boer, K. Hori, H. Ooguri and Y. Oz,
{\it ``Branes and Dynamical Supersymmetry Breaking'',}
\npb{522}{1998}{20,}
\bb{9801060.}}
\lref\rdhook{J. de Boer, K. Hori, H. Ooguri and Y. Oz,
{\it ``Kahler Potential and Higher Derivative Terms from M Theory 
Fivebrane'',}
\npb{518}{1998}{173,}
\bb{9711143.}}
\lref\op{Y. Oz and A. Pasquinucci, work in progress.}
 
\lref\rshifk{A. Kovner,  M. Shifman and A. Smilga,
{\it ``Domain Walls in Supersymmetric Yang-Mills Theories'',}
\prd{56}{1997}{7978,} 
\bb{9706089.}}
 
\lref\rwittheta{E. Witten, 
{\it ``Large N Chiral Dynamics",} 
{\it Ann. Phys.} 128 (1980) 363.} 
\lref\rdashen{R.F. Dashen, 
\prd{3}{1971}{1879.}}
\lref\rwittlargen{E. Witten, 
{\it ``Current Algebra Theorems for the $U(1)$ `Goldstone Boson' '',}
\npb{156}{1979}{269.}}

\lref\revhsub{N. Evans, S.D.H. Hsu and M.Schwetz, 
{\it ``Phase Transitions in Softly Broken N=2 SQCD at nonzero Theta Angle'',} 
\npb{484}{1997}{124,} 
\bb{9608135.}}
 
\lref\rmqcd{A. Hanany, M.J. Strassler and A. Zaffaroni, 
{\it ``Confinement and Strings in MQCD",} 
\npb{513}{1998}{87,} 
\bb{9707244.}} 
\lref\revans{N. Evans and M. Schwetz,
{\it ``The Field Theory of Non-Supersymmetric Brane Configurations'',}
\npb{522}{1998}{69,}
\bb{9708122.}}
\lref\rlastev{N. Evans, 
{\it ``Quark Condensates in Non-supersymmetric MQCD'',}
\bb{9801159.}}
\lref\rpelc{S. Elitzur, O. Pelc and E. Rabinovici,
{\it ``Aspects of Confinement and Screening in M-Theory'',} 
\bb{9805123.}}
\lref\rcobi{Y. Kinar, E. Schreiber and J. Sonnenschein,
{\it ``Precision `Measurements' of the $Q \bar{Q}$ Potential in MQCD'',}
\bb{9809133.}}

%%%%%%%%%%%%
%% either    refs.\ \refs{\rwinst,\rdoug,\rvafi,\rdm}
%% or        ref.\ \rdoug\
%% or        ref.\ \refs\rvafi\
%%
%% eqs:      \eqn\metric{ x = y}
%% or        \eqn\conds{\eqalign{ & x=y \cr & z=t \cr}}
%%
%% also      \newsec{title}
%%%%%%%%%%%%

%%%%%%%%TEXT%%%%%%%%%%%%%%%%%%%%%%%%%%%%%%%%%%%%%%%%%%%%%%%%%%%%%%%%%
%%%%%%%%%%%%%%%%%%%%%%%%%%%%%%%%%%%%%%%%%%%%%%%%%%%%%%%%%%%%%%%%%%%%%
%%%%%%%%%%%%%%%%%%          title page       %%%%%%%%%%%%%%%%%%%%%%%%%
%%%%%%%%%%%%%%%%%%%%%%%%%%%%%%%%%%%%%%%%%%%%%%%%%%%%%%%%%%%%%%%%%%%%%%

\line{\hfill CERN-TH/98-299}
\line{\hfill KUL-TF-98/35}
\line{\hfill {\tt hep-th/9809173}}
\vskip 1.2cm

\Title{\vbox{\baselineskip 12pt\hbox{}
 }}
{\vbox {\centerline{Branes and Theta Dependence}
%\vskip10pt
%\centerline{    }
}}

\vskip0.6cm

\centerline{$\quad$ {\caps Yaron  Oz~$^1$ and Andrea Pasquinucci~$^{1,2}$
 }}
\vskip0.8cm

\centerline{{\sl $^1$ Theory Division, CERN}}
\centerline{{\sl 1211 Geneva 23, Switzerland}}

\vskip0.3cm

\centerline{{\sl  $^2$ Instituut voor theoretische fysica, K.U.\
Leuven}}
\centerline{{\sl  Celestijnenlaan 200D }}
\centerline{{\sl  B-3001 Leuven, Belgium  }}

\vskip 1.0in

\noindent{\bf Abstract:}
We use  the fivebrane of M theory to study the $\theta$ dependence
of four dimensional $SU(N_c)$ super Yang-Mills
and super QCD
softly broken by a gaugino mass.
We compute the  energy of the vacuum 
 in the supergravity approximation.
The results obtained 
are in qualitative agreement with field theory.
We also study the $\theta$ dependence of the QCD string tension 
via the fivebrane. 
 
%
%\vskip8pt\hrule\vskip8pt
%
%\centerline{AREA VERSION 03}
%

%%%%%%%%%%%%%%%%%%%%%%%%%%%%%%%%%%%%%%%%%%%%%%%%%%%%%%%%%%%%%%%%%%%%%%

\Date{September 1998}

%\draft

%%%%%%%%%%%%%%%%%%%%%%%%%%%%%%%%%%%%%%%%%%%%%%%%%%%%%%%%%%%%%%%%%%%%%%%%%%
%%%%%%%%%%%%                text begins                        %%%%%%%%%%%
%%%%%%%%%%%%%%%%%%%%%%%%%%%%%%%%%%%%%%%%%%%%%%%%%%%%%%%%%%%%%%%%%%%%%%%%%%
%
%
\newsec{Introduction}

The $\theta$ dependence of four dimensional gauge theories encodes 
non trivial information of the dynamics.
A study of this dependence in
 asymptotically free gauge theories requires an appropriate
effective low energy description
of the system, which is difficult to obtain.
 Recently, Witten \ \refs\rwittentheta\ 
studied the $\theta$ dependence of the vacuum energy of 
large $N_c$ four dimensional gauge theory using the conjectured
supergravity/gauge theory correspondence \ \refs{\maldacena,\wittenads}.

Another approach to study four dimensional gauge theories
is via the fivebrane of M theory.
When the fivebrane is wrapping a holomorphic curve, the effective
low energy four dimensional gauge theory is supersymmetric.
Using this description, many
 holomorphic properties of these theories have been derived \ \refs\givkut.
In this paper we will use the fivebrane of M theory 
to  study the theta dependence 
of softly broken supersymmetric four 
dimensional super Yang-Mills (SYM) and  super QCD (SQCD).~\foot{For related 
works see refs.\ \refs{\rthetus,\revhsub,\rlastev}.}

The paper is organized as follows:
In the section 2 we will consider SYM softly broken by a gaugino mass.
We will first review the field theory computation of the vacuum energy
and the physics encoded in the $\theta$ dependence.
We will then compute the same quantity using the brane description.
The two results agree qualitatively and have the same $\theta$ dependence.
We will compute the QCD string tension via the fivebrane.
Here we will have no field theory result to compare with.
We will close the section by discussing the large $N_c$ limit.
In section 3 we will consider SQCD with massive matter
in the fundamental representation of the gauge group,
softly broken by a gaugino mass.
We will compute the vacuum energy in field theory and using 
the fivebrane description. Again we obtain qualitative agreement.
Section 4 is devoted to a discussion and comments on the
 decay rate of the false
vacuum.

\newsec{Softly Broken SYM}

\subsec{Effective Field Theory of SYM}
 
Consider the Veneziano-Yankielowicz (VY) effective action 
\ \refs\VY\ for 
$SU(N_c)$ super 
Yang-Mills softly broken by a gaugino mass. We will follow in the discussion
the conventions of ref.\ \refs\rmasiero. Denote by $S$ the VY superfield
and by $\varpi$ its first component. 
The composite superfield $S$ is defined by 
$S = (g^2/ 32 \pi) W^a_\alpha W^{a,\alpha}, \alpha=1,2, a = 1,...,N_c^2-1$.  
$\varpi = (g^2/ 32 \pi) \lambda^a_\alpha \lambda^{a,\alpha}$ where $\lambda^{a,\alpha}$
denotes the gaugino. The effective
VY action is
\eqn\vyaction{{\cal L} = {9\over \alpha} (\bar{S}S)^{1/3}_D + \left[
\left( S\left( N_c \log{S\over \vert\Lambda_1\vert^3} -N_c -i\theta\right)
\right)_F + h.c.\right] + [m_\lambda S_{\theta=0} + h.c. ], }
where the (complex) gaugino mass $m_\lambda$ is the renormalization group 
invariant supersymmetry breaking parameter. The theta angle is defined 
 by
\eqn\eqtheta{\theta = N_c {\rm arg}((\Lambda_1)^3),}
where $\Lambda_1$ is the dynamically generated scale. 
The scalar potential resulting from the VY action is given by 
\eqn\vymsv{ V= \alpha (\varpi\bar\varpi)^{2/3} \left\vert N_c \log
{\varpi\over \vert\Lambda_1\vert^3} -i\theta\right\vert^2 
-(m_\lambda\varpi + \bar{m}_\lambda\bar\varpi)\ ,}
and the $N_c$ supersymmetric  minima  (when $m_{\lambda}=0$) are located at
\eqn\vymmin{ \langle \varpi\rangle_{susy,k} = \vert\Lambda_1\vert^3 \exp 
\left( i {\theta + 2\pi k\over N_c}\right),\qquad k=0,\cdots,N_c-1.} 
The degeneracy corresponds to the spontaneous breaking of the
non anomalous $Z_{2N_c}$ discrete subgroup of $U(1)_R$ to $Z_2$.

We assume that the gaugino mass softly breaks the supersymmetry, 
$\vert m_\lambda\vert << \vert\Lambda_1\vert$. In this regime we are
interested in computing quantities only to first order in $m_\lambda$.
Consider a generic scalar potential of the form
\eqn\vmgenv{V(\varphi_i)=V_{susy}(\varphi_i) - 
(m_\lambda\varpi + \bar{m}_\lambda\bar\varpi)\ .}
When $m_\lambda=0$ the vacuum energy vanishes and
\eqn\vmgenvac{V_{susy}(\langle\varphi_i\rangle_{susy})=0\ , 
\qquad\qquad\qquad {\partial V_{susy}(\langle\varphi_i\rangle_{susy})
\over \partial \varphi_j} =0. }
When supersymmetry is softly broken we look for a vacuum configuration
of the form
\eqn\vmvacb{\langle\varphi_i\rangle = \langle\varphi_i\rangle_{susy}
+ m_\lambda \langle\varphi_i\rangle_{1} + O((m_\lambda)^2)\ . }
Substituting this in eq.\ \vmgenv\ and Taylor expanding for small $m_\lambda$
around $\langle\varphi_i\rangle_{susy}$ we obtain
\eqn\vmgenvb{V(\langle\varphi_i\rangle) =  - 
(m_\lambda\langle\varpi\rangle_{susy} + 
\bar{m}_\lambda\langle\bar\varpi\rangle_{susy}) +O((m_\lambda)^2).}
Thus to compute, to first order in $m_{\lambda}$
 the vacuum energy (per unit volume) we need to know
only $\langle\varpi\rangle_{susy}$. In the case at hand this is given
by eq.\ \vymmin\ and we obtain  \refs\rmasiero
\eqn\vymener{ E_k(\theta,N_c,m_\lambda,\Lambda_1) \simeq 
- 2 \vert m_\lambda\vert \, \vert \Lambda_1\vert^3 
\cos \left( {\theta_p +2\pi k\over N_c}\right),}
where
\eqn\vymthetap{\theta_p = \theta + N_c {\rm arg} (m_\lambda)\ ,}
is the physical $\theta$ angle.

The gaugino mass term breaks explicitly the $Z_{2N_c}$ symmetry and 
shifts the $N_c$ local minima. 
The true ground state is given by the value of $k$ that 
minimizes eq.\ \vymener. 
When $-\pi < \theta_p < \pi$ the true vacuum is at $k=0$. At $\theta=\pi$ 
the $k=0$ and $k=1$ levels cross, and the $k=1$ vacuum is the true ground state
up to $\theta_p=3\pi$ where it crosses the $k=2$ vacuum, and the $k=2$
vacuum becomes the true ground state up to $k=3$. This pattern continues until
$\theta_p = (2N_c-1)\pi$ where we have again the $k=0$ vacuum.
Thus $\theta_p$ periodicity is $2\pi N_c$
while the physics is periodic with period $2\pi$. When 
$\theta_p=(2n+1)\pi$, two vacua have the same energy and CP is spontaneously
broken \refs{\rdashen,\rwittheta}.

\subsec{Brane Computation of Vacuum Energy}

Consider the M5 brane with its $6d$
world-volume  being $R^4 \times \Sigma$ embedded in $R^4 \times R^6 \times S^1$.
$\Sigma$ is a real two dimensional space.
We denote by $x^0,...,x^3$ and $x^4,...,x^9$  the coordinates
on $R^4$ and $R^6$ respectively, and by $x^{10}$ the circle coordinate.
The radius of the circle is taken to be $R$.
Introduce the complex coordinates~\foot{In our notations, $x^4,x^5,x^7,x^8$ 
and $x^9$  have dimension of mass, $x^0,...,x^3$, $x^6$ and $x^{10}$ have
dimension of length.}
$v=x^4+ix^5$, $w=x^8+ix^9$, $t=\exp[-(x^6+ix^{10})/R]$.

Let $\Sigma$ be  given by 
\eqn\curvezero{\eqalign{ & v= z + {\epsilon\over\bar{z}},
\qquad\qquad\quad  w= {\zeta\over z}, \cr & t = z^{N_c}, \qquad\qquad\qquad
x^7 = 2\sqrt\epsilon \log\left\vert{z\over\Lambda_1}\right\vert\ , \cr}}
while the $R^4$ world-volume coordinates are $x^0,..,x^3$. 
The flat eleven dimensional background metric is given by 
\eqn\eqmetric{ (ds)^2 = \sum_{m,n=0}^3 \eta_{mn}dx^mdx^n+
2G_{v\bar{v}}dvd\bar{v} + 
2G_{w\bar{w}}dwd\bar{w} + 2G_{w\bar{w}}dwd\bar{w} + 
G_{77}dx^7dx^7, }
where
\eqn\eqgg{\eqalign{G_{v\bar{v}} & = { (l_p)^6 \over 2 (2\pi R)^2}
\qquad\qquad\qquad\  G_{w\bar{w}}  = { (l_p R)^2 \over 2} \cr
G_{t\bar{t}} & = { (R)^2 \over 2 \vert t\vert^2}
\qquad\qquad\qquad\qquad  G_{77} = { (l_p)^6 \over (2\pi R)^2}. \cr}}

The effective low energy four dimensional gauge theory in $x^0,...,x^3$ 
described by this system is $SU(N_c)$ Yang-Mills 
with softly broken supersymmetry \ \refs{\rwitqcd,\rthetus}.
The real parameter $\epsilon$ describes the soft breaking of 
supersymmetry:~\foot{In the supergravity approximation, 
the parameter $\epsilon$ is real and
positive. It is possible that beyond  
the supergravity approximation (or for
hard supersymmetry breaking) it 
 acquires an imaginary part which can signal the 
disappearance of a metastable vacuum for large values of $\theta_p$, as
discussed in ref.\ \rthetus.}
for $\epsilon=0$ we have a holomorphic
 curve and a  realization of 
$N=1$ supersymmetric Yang-Mills \ \refs{\rmberk,\rwitqcd} 
while for $\epsilon\neq 0$
we get a  soft breaking of supersymmetry by gaugino mass.
When  $\epsilon=0$
\eqn\zzzeta{\zeta_{susy} = C_\zeta {(l_p)^2\over R} (\Lambda_1)^3,}
with $C_{\zeta}$ a real dimensionless constant which a priori can 
depend on $N_c$.
When  $\epsilon=0$ \curvezero\  describes $N_c$
supersymmetric curves, which correspond to the $N_c$ supersymmetric
vacua of SYM, given by the different values of the phase of $\Lambda_1$ :
\eqn\bzetak{\zeta_k = \vert \zeta_{susy}\vert\, 
\exp(i{\theta + 2\pi k\over N_c}), } 
where
\eqn\eqthetaequiv{\theta\equiv N_c\,{\rm arg}(\zeta_{susy})\vert_{k=0} = 
 N_c\,{\rm arg}((\Lambda_1)^3)\vert_{k=0},\qquad k=0,\cdots,N_c-1.}

The parameter $\epsilon$ is a function of the
gaugino mass $m_\lambda$. It vanishes when $m_\lambda=0$. 
When supersymmetry is broken, also $\zeta$ can acquire a dependence 
on $m_\lambda$ which we can parametrize as follows
\eqn\parszero{\zeta(m_\lambda) = \zeta_{susy}( 1 + f_\zeta(m_\lambda) ),}
where $f_\zeta$ vanishes when $m_\lambda$ does.
The charge assignments for the physical parameters are given by \refs\rmberk\
\eqn\mcharge{\vbox{\tabskip=0pt \offinterlineskip 
\halign to 110pt{ \strut#&  \hfil #& \vrule #\tabskip=1em plus2em &
\hfil #\hfil & \hfil #\hfil \cr 
      &             &     &$U(1)_{45}$  & $U(1)_{89}$ \cr
      \noalign{\hrule}  
      & $m_\lambda$ &     &  $-2$      &    $-2$     \cr
      &$(\Lambda_1)^3\,$& &   2        &     2     \cr
      & $R$          &    &    0       &     0     \cr}}}
where  $U(1)_{45}$ and  $U(1)_{89}$ denote the rotations in $x^4,x^5$ and $x^8,x^9$
respectively.  
Using  dimensional analysis,
the charges \ \mcharge\ and the fact that $\epsilon$ and $f_\zeta$ 
do not carry any charge,
we obtain the following expansions to first order in $m_\lambda$ 
\eqn\parfzero{\eqalign{ & \epsilon(m_\lambda) = C_\epsilon m_\lambda
(R)^2 (\Lambda_1)^3 + c.c. + O((m_\lambda)^2) =
 2 C_\epsilon \cos(\theta_{p,k}/N_c) \vert m_\lambda\vert (R)^2
\vert\Lambda_1\vert^3 + \cdots\cr
& f_\zeta(m_\lambda)  = C_{f_\zeta} m_\lambda (R)^4 (\Lambda_1)^3 +
O((m_\lambda)^2),  \cr}}
where
\eqn\zthetad{\theta_{p,k} =\theta_p  +2\pi k =  \theta + 
N_c\,{\rm arg}(m_\lambda)+2\pi k\ . }
A priori the real constants $C_\epsilon$ and $C_{f_\zeta}$ can depend 
on $N_c$.

The classical Nambu-Goto action of the M5 brane suggests that the area of
$\Sigma$ plays a role of a potential energy \refs\dhoo.
The energy  (per unit volume) of the configuration is related to the area 
of the five-brane by
E~=~Area/$(l_p)^6$.
The area of $\Sigma$ is given by
\eqn\areazero{Area = \int_{\Sigma} {\rm d}^{\,2}z \left\{ G_{i\bar{j}}
(\partial x^i \bar\partial
\bar{x}^{\bar{j}} + \bar\partial x^i \partial \bar{x}^{\bar{j}})
+ G_{77} \partial x^7 \bar\partial x^7\right\} }
where $(i,j)\in\{v,w,t\}$. However, since $\Sigma$ is non compact
the area is infinite. We therefore have to define a notion of a 
regularized area. This will be done  by subtracting the
infinite area of the holomorphic (supersymmetric) curve
\eqn\areaone{Area = Area\vert_\epsilon -
Area\vert_{\epsilon=0}.}
With this definition 
the $Area$ of a holomorphic curve vanishes giving zero energy for
the supersymmetric vacua. Computing \ \areaone\ to first order
in $m_\lambda$ we  obtain for the vacuum energy 
\eqn\areafour{E_k(\theta,N_c,m_\lambda,\Lambda_1)  = {\rm V}_1\,
\vert m_\lambda\vert \vert\Lambda_1\vert^3\,
\cos\left({\theta_{p}+ 2\pi k\over N_c}\right) +
O(\vert m_\lambda\vert^2),}
where $k=0,...,N_c-1$ and $V_1$ is given by
\eqn\areafourv{ {\rm V}_1 =  \int_0^{+\infty} x{\rm d}x 
\left\{ {C_\epsilon\over \pi}
{1\over x^2} + (C_\zeta)^2 C_{f\zeta}\vert\Lambda_1\vert^4
(R)^4 {2\pi \over x^4} \right\} ,}
and $x=\vert z\vert/\vert\Lambda_1\vert$ is a dimensionless parameter.

The true vacuum is obtained by minimization of \areafour\ with respect to $k$.
The result  \areafour\ agrees with the general analysis of ref.\ \rthetus\ 
and exhibits the same qualitative behaviour and in particular the same 
$\theta$ dependence as the field theory result \ \vymener.
The overall coefficient is different. In the brane 
computation it depends explicitly on $R$, and furthermore 
${\rm V}_1$ is divergent and requires 
a suitable
regularization. However, it is by now clear that the eleven dimensional 
supergravity description that we are using is not capable of a reliable 
computation of the numerical coefficients \refs{\dhoo,\rdhook}.
Note that the result \ \areafour\ interpolates smoothly between the SYM
and ordinary Yang-Mills.

\subsec{Brane computation of QCD string tension}

A QCD string that carries $l~ mod~ N_c$ units of flux is realized in the 
brane framework as a curve connecting 
two points on $\Sigma$,
$z_0$ and $z_0 \exp[2\pi i l/N_c]$  
\refs{\rwitqcd} \foot{For other work on the QCD string 
in the brane picture see \refs{\rmqcd,\revans,\rpelc,\rcobi}.}. 
The QCD string tension is given
by the minimal length between two such points. The distance
between the two points on $\Sigma$  is given by
\eqn\disdef{ dist(z_0) = \sqrt{2G_{v\bar{v}} \vert \Delta v\vert^2 +
2G_{w\bar{w}} \vert \Delta w\vert^2 +
2G_{t\bar{t}} \vert \Delta t\vert^2 +
G_{77} (\Delta x_7)^2 }\ .}
Computing the distance and taking the
minimum with respect to $z_0$ we obtain the tension of the $l$-string 
as $T_l=dist(z_{0,min})/(l_p)^3$:
\eqn\strten{T_l = {2\over l_p} \sqrt{{\vert\zeta\vert\over \pi}} 
\vert\sin(l\pi/N_c)\vert 
\left(1 + {(l_p)^2 \pi\over \vert\zeta\vert (2\pi R)^2}
\epsilon \right) + O(\epsilon^2). }
To first order in $m_\lambda$ we get
\eqn\strtenm{\eqalign{T_l =& {2\over l_p}
\sqrt{{\vert\zeta_{susy}\vert\over \pi}}  \vert\sin(l\pi/N_c)\vert  
\Big(1  +\cr
&\quad \left. {1\over 2} \vert m_\lambda\vert \,\vert\Lambda_1\vert^3 (R)^2 
\left[ C_{f_\zeta} (R)^2 + {1\over \pi} \left({l_p\over R}\right)^2
{ C_\epsilon \cos(\theta_{p,k}/N_c) \over \vert\zeta_{susy}\vert } \right]
\right) + O(\vert m_\lambda\vert^2)\ .\cr }}
Unlike the vacuum energy, now we have no field theory result to compare with.

\subsec{Large $N_c$ limit}

Consider now  
the large $N_c$ limit.  
Since we do not know 
 the behaviour in the large-$N_c$ limit of the higher order
terms in the expansion in $m_\lambda$, we will make the assumption
that these terms  are not of higher order in $N_c$.
The analysis of \ \refs{\rwitqcd}\ in the $N=1$ supersymmetric case gave  
$R\sim O(1/N_c)$ and $\zeta_{susy}\sim O((N_c)^2))$. Together with 
\ \zzzeta\ 
we get $C_\zeta \sim O(1)$.
Since we do not know the large $N_c$ behaviour 
of $\zeta$ when supersymmetry is broken we assume that
$\zeta \sim \zeta_{susy} \sim O((N_c)^2)$.
This implies by \ \parfzero\ that $C_{f_\zeta} \leq O((N_c)^2)$.

One expects for the vacuum energy that \refs{\rwittlargen,\rwittheta}
\eqn\areafourl{E_k(\theta)  = C \left(\theta_{p}+ 2\pi k \right)^2 + O(1/N_c),}
with the coefficient $C$ being  a constant independent of $N_c$.
Since $m_{\lambda}$ and $(\Lambda_1)^3$ are
proportional to $N_c$ in the large-$N_c$ limit, eq. \ \areafour\ and 
\areafourv\ imply that ${\rm V}_1 \sim O(1)$.
Assuming a suitable regularization of $V_1$ it implies 
$C_\epsilon \sim  O(1)$ while the second term 
in ${\rm V}_1$ vanishes in the limit.
Note also  that  $\epsilon(m_\lambda) \sim O(1)$. 

The implication of the above discussion
to the large $N_c$ behaviour of the QCD string tension eq.\ \strtenm\ is
that the $m_\lambda$ term within the square brackets 
of this equation is $O(1)$.

\newsec{Softly Broken SQCD}

\subsec{Effective Field Theory of SQCD}

Consider super-QCD with gauge group 
$SU(N_c)$ and $N_f$ flavours. For  $0<N_f<N_c$ we consider the
Taylor-Veneziano-Yankielowicz (TVY) effective action 
\refs\TVY\ 
where the mesonic matter superfields $T$ all have equal (complex) mass $m_f$ 
and supersymmetry is softly broken
by a gaugino mass term
\eqn\tvyaction{\eqalign{{\cal L} = &{9\over \alpha} (\bar{S}S)^{1/3}_D + 
{1\over \gamma} (\bar{T}^i_j T^j_i)(\bar{S}S)^{-1/3}_D
+\left[\left( S\left( \log{S^{N_c-N_f} \det T \over 
\vert\Lambda_1\vert^{3N_c-N_f}} -(N_c-N_f) -i\theta'\right)
\right)_F + h.c.\right] \cr &+  [{\rm Tr}(m_f T)_F + h.c.] + 
[m_\lambda S_{\theta=0} + h.c. ]\ ,\cr}}
where 
\eqn\thetatvy{\theta' = \theta\left(1-{N_f\over 3N_c}\right) \ . }
In this case
\eqn\tvymvac{ \langle \varpi\rangle_{susy,k} = \vert\Lambda_1\vert^{3-N_f/N_c}
\vert m_f \vert^{N_f/N_c} 
\exp \left( i {\theta' + N_f{\rm arg}(m_f) + 2\pi k\over N_c}\right),} 
with $k=0,\cdots,N_c-1$, and the vacuum energy (per unit volume) is given by
\eqn\tvymener{ E_k(\theta,N_c,N_f,m_\lambda,m_f,\Lambda_1) \simeq 
- 2 \vert m_\lambda\vert \, \vert \Lambda_1\vert^{3-N_f/N_c}
\vert m_f\vert^{N_f/N_c} 
\cos \left( {\theta_p +2\pi k\over N_c}\right),}
where
\eqn\tvymthetap{\theta_p = \theta' + N_f{\rm arg}(m_f) + N_c {\rm arg} (m_\lambda)\ .}

When $N_F \geq N_c$ extra baryonic superfields must be added to the TVY
effective action. Notice that in the regime we are considering the
energy of the vacuum configuration depends only on the supersymmetric
expectation value of $\varpi$ \tvymvac\ (see the discussion in section 2.1). 
We therefore expect  that
the vacuum energy is still given by
\tvymener. As we will see this is indeed the result of the brane computation.

\subsec{Brane Computation of Vacuum Energy}

We will now add massive matter in the fundamental
representation of the gauge group. 
For simplicity we assume that all $N_f$ matter multiplets have the same
mass $m_f$. $\Sigma$ is given by
\eqn\curveone{\eqalign{ & v= z + {\epsilon\over\bar{z}}
\qquad\qquad\qquad\quad  w= {\zeta\over z} \cr
& t = {z^{N_c} \over (z-z_-)^{N_f}} \qquad\qquad\quad
x^7 = 2\sqrt\epsilon \log\left\vert{z\over\Lambda_1}\right\vert\ .\cr}}
In the supersymmetric case ($\epsilon=0$) the parameters of the curve
are related to the physical parameters by
\eqn\parsone{\eqalign{& \zeta_{susy} = C_\zeta {(l_p)^2\over R}
(\Lambda_1)^{3-N_f/N_c} (m_f)^{N_f/N_c} \cr
& z_{-,susy} = -m_f\ .\cr}}
When $\epsilon\neq 0$ supersymmetry is softly broken. As discussed in
ref.\ \refs\rbrokus, this soft breaking corresponds in field theory to
introducing only a gaugino mass $m_{\lambda}$.
The parameters in $\Sigma$ can be written as
\eqn\parstwo{\eqalign{& \zeta(m_\lambda) = \zeta_{susy} 
(1 + f_\zeta (m_\lambda))\cr
& z_{-}(m_\lambda) = -m_f(1+f_{z_-}(m_\lambda))\ ,\cr}}
where $f_\zeta(m_\lambda)$, $f_{z_-}(m_\lambda)$ and $\epsilon(m_\lambda)$
vanish when $m_\lambda$ does. 

Using the  charge assignments \refs\rmberk\
\eqn\mcharge{\vbox{\tabskip=0pt \offinterlineskip 
\halign to 160pt{ \strut#&  \hfil #& \vrule #\tabskip=1em plus2em &
\hfil #\hfil & \hfil #\hfil \cr 
      &                    &  & $U(1)_{45}$ & $U(1)_{89}$ \cr
      \noalign{\hrule}  
      & $m_\lambda$        & &  $-2$        &    $-2$     \cr
      & $m_f$              & &  $2$         &    $0$     \cr
&$(\Lambda_1)^{3N_c-N_f}\,$& & $2N_c-2N_f$  &  $2N_c$     \cr
      & $R$                & &    0         &     0     \cr}}}
and the fact that $\epsilon$, $f_{z_-}$ and $f_\zeta$ do not carry any charge,
we obtain the following expansions to first order in $m_\lambda$ 
\eqn\parsthree{\eqalign{&\epsilon(m_\lambda) = 2C_\epsilon
 (R)^2 \vert m_\lambda\vert\, \vert\Lambda_1\vert^{3-N_f/N_c} 
\vert m_f\vert^{N_f/N_c} 
\cos\left({\theta_p +2\pi k\over N_c}\right)+
O(\vert m_\lambda\vert^2)\cr
& f_\zeta(m_\lambda) = C_{f\zeta} m_\lambda(R)^4 (\Lambda_1)^{3-N_f/N_c}
(m_f)^{N_f/N_c} + O(( m_\lambda)^2)\cr
& f_{z_-}(m_\lambda) = C_{fz_-}m_\lambda (R)^4
(\Lambda_1)^{3-N_f/N_c} (m_f)^{N_f/N_c} + O((m_\lambda)^2)\ ,\cr}}
where 
\eqn\thetapbtwo{\theta_p = \left(1-{N_f\over 3N_c}\right)\theta + 
N_c {\rm arg}(m_\lambda) + N_f {\rm arg}(m_f)\ .} 
The computation of the energy (per unit volume) is  done as in 
softly broken SYM. We get to order $O(m_{\lambda})$
\eqn\newathree{E_k(\theta,N_c,N_f,m_\lambda,m_f,\Lambda_1) = {\rm V}_2
\vert m_\lambda\vert \vert\Lambda_1\vert^{3-N_f/N_c}
\vert m_f\vert^{N_f/N_c} 
\cos\left({\theta_p +2\pi k\over N_c}\right),}
with $k=0,\cdots,N_c-1$ and
\eqn\newathreev{\eqalign{ {\rm V}_2 =& \int_0^{+\infty} x{\rm d}x \left\{
{1\over \pi}{C_\epsilon\over x^2} + C_{f\zeta} (C_\zeta)^2
(R)^4 \vert\Lambda_1\vert^{4-2N_f/N_c} \vert m_f\vert^{2N_f/N_c}
{2\pi \over x^4}\right. \cr
&\qquad\qquad\quad - \left. C_{fz_-} (N_f)^2
\left({R\over l_p}\right)^6 {\vert\Lambda_1\vert\over\vert m_f\vert}
{2\pi \over (1-x^2)^2} \right\}.\cr }.}
Comparing \newathree\ and the field theory results
\tvymener\ we see that  they have the
same qualitative behaviour and $\theta$ dependence.

Taking the limit $m_f\rightarrow\infty$ while keeping
$(\widetilde\Lambda_1)^3 = (\Lambda_1)^{3-N_f/N_c} (m_f)^{N_f/N_c}$ fixed
(with $N_f <3N_c$) and rescaling appropriately the integration variables,
one decouples all the matter fields and obtains the results  of the
previous section, as expected.

\newsec{Discussion and Conclusions}

We computed, in the supergravity approximation,
the vacuum energy of $N=1$ SYM and SQCD with massive matter,
softly broken by a gaugino mass using their
realization via the fivebrane of M theory.
The results we obtained matched qualitatively the  field theory expectations.
In particular the $\theta$ dependence and the physics associated with
it were  in agreement.
This supports the hope that the field and brane theories are in 
the same universality class.
Our analysis is valid for finite $N_c$ with soft supersymmmetry
breaking parameter
$m_{\Lambda}$.
Note in comparison, that in
\ \refs\rwittentheta\ the analysis was carried out for large $N_c$ and
hard supersymmetry breaking.

We also computed the QCD string tension via the fivebrane realization.
It would be interesting to perform a field theory computation to see if there
is a qualitative agreement with the brane result of the $\theta$
angle dependence.

Another interesting issue is the decay rate of the false vacuum.
In the semiclassical approximation it is given by
\eqn\vmdecay{ {\Gamma \over Vol} \sim {1 \over \vert\Lambda_1\vert^4}
\exp\left[-{27\over 2}\pi^2 {(T_D)^4\over \vert \Delta E\vert^3}\right],  }
where $T_D$ is the tension of the domain wall
separating the two vacua and
$\Delta E$ is the difference of their energies.
Since
$\Delta E$ is of order $m_\lambda$ while
\eqn\tdeq{ (T_D)^4 = (T_{D,susy})^4 + 4 m_\lambda \tilde{T}_{D,broken} (T_{D,susy})^3
+ O((m_\lambda)^2),} the dominant term in the decay rate
for small $m_\lambda$ is $T_{D,susy}$ \refs\rshifk. 
Therefor the decay rate predicted by field theory and the brane theory 
agree qualitatively
and have same $\theta$ dependence. 
It would also be interesting to see whether the 
agreement continues to hold when including the $O(m_{\lambda})$
correction to the domain wall tension.
In this case, both the field theory and the brane theory results are not known.
In order to get the decay rate \ \vmdecay\ purely from brane 
considerations we need to use the realization of the domain wall
connecting two adjacent vacua as a three manifold interpolating between
the two dimensional real manifolds $\Sigma, \Sigma'$ \ \refs\op.

\vskip 0.9cm
{\bf Acknowledgements}
\vskip 0.1cm
We thank M. Shifman for useful discussions.
This work is partially supported by the European Commission TMR programme
ERBFMRX-CT96-0045 in which A.P.\ is associated to the Institute for
Theoretical Physics, K.U.\ Leuven. A.P.\ would like to thank CERN for its
hospitality while part of this work was carried out.

\listrefs
\vfill
\eject
%\bye
\end